\documentclass[journal,onecolumn]{IEEEtran}
\IEEEoverridecommandlockouts
\usepackage{fix-cm}
\usepackage{cite}
\usepackage{amsmath,amssymb,amsfonts}
\usepackage[ruled,vlined]{algorithm2e}
\usepackage{algorithmic}
\usepackage{graphicx,color}
\usepackage{comment}
\usepackage{xcolor}
\usepackage{multirow}
\usepackage{url}
\usepackage{textcomp}
\usepackage{adjustbox}
\usepackage{balance}
\usepackage{booktabs}
\usepackage{hyperref}

\begin{document}

\title{Decentralized Orchestration Architecture for Fluid Computing: A Secure Distributed AI Use Case\\
}

\author{
    \IEEEauthorblockN{Diego Cajaraville-Aboy\textsuperscript{*}, Ana Fernández-Vilas, Rebeca P. Díaz-Redondo, Manuel Fernández-Veiga, Pablo Picallo-López}\\
    \IEEEauthorblockA{atlanTTic Research Center -- ICLAB -- Universidade de Vigo, Vigo, 36310, Spain\\
    \{dcajaraville,avilas,rebeca,mveiga,iclab\}@det.uvigo.es}\\
    \textsuperscript{*}{Corresponding author: dcajaraville@det.uvigo.es}%
    \thanks{This version of the article has been accepted for publication in Computer Networks after peer review, but is not the Version of Record and may not reflect final copy-editing, formatting, pagination, or corrections. The Version of Record is available online at: \url{https://doi.org/10.1016/j.comnet.2026.112369}. © 2026 The Authors. This manuscript version is made available under the CC BY-NC-ND 4.0 license.}%
}

\maketitle

\begin{abstract}
Distributed AI and IoT applications increasingly execute across heterogeneous resources spanning end devices, edge/fog infrastructure, and cloud platforms, often under different administrative domains. Fluid Computing has emerged as a promising paradigm for enhancing massive resource management across the computing continuum by treating such resources as a unified fabric, enabling optimal service-agnostic deployments driven by application requirements. However, existing solutions remain largely centralized and often do not explicitly address multi-domain considerations. This paper proposes an agnostic multi-domain orchestration architecture for fluid computing environments. The orchestration plane enables decentralized coordination among domains that maintain local autonomy while jointly realizing intent-based deployment requests from tenants, ensuring end-to-end placement and execution. To this end, the architecture elevates domain-side control services as first-class capabilities to support application-level enhancement at runtime. As a representative proof of concept, we instantiate the architecture through a distributed AI use case; specifically, we consider a multi-domain Decentralized Federated Learning (DFL) deployment under Byzantine threats. Under this setting, we leverage domain-side capabilities to enhance Byzantine security by introducing FU-HST, an SDN-enabled multi-domain anomaly detection mechanism that complements Byzantine-robust aggregation. We validate the use-case workflow via simulation in single- and multi-domain settings, evaluating anomaly detection, DFL performance, and computation/communication overhead.
\end{abstract}


\section{Introduction}
\label{sec:introduction}

\IEEEPARstart{I}{n} recent years, the emergence of paradigms such as the Internet of Things (IoT) or techniques such as artificial intelligence (AI) has led to unprecedented growth in intelligent service deployments in various fields. The digital infrastructure has evolved rapidly to support today's complex software ecosystem and numerous data-driven applications~\cite{gill2024modern, drechsler2022digital}. This evolution has enabled the deployment of distributed AI services on a global scale and closer to end users, thus empowering time-constrained critical applications and preserving user privacy. To accommodate these trends, current networks must evolve towards AI-native architectures that integrate computing capabilities into IoT devices and wireless nodes~\cite{alberti2024disruptive}, as well as enable device-to-device (D2D) communications, given the emergence of standards such as 5G-Advanced and the upcoming 6G networks~\cite{chen20235g}. Furthermore, these distributed applications still face significant challenges in managing vast amounts of data while ensuring robust security and privacy mechanisms against potential adversarial attacks on these distributed computing environments~\cite{nair2023robust}.

\begin{figure*}[tbp]
    \centering
    \includegraphics[width=1.0\linewidth]{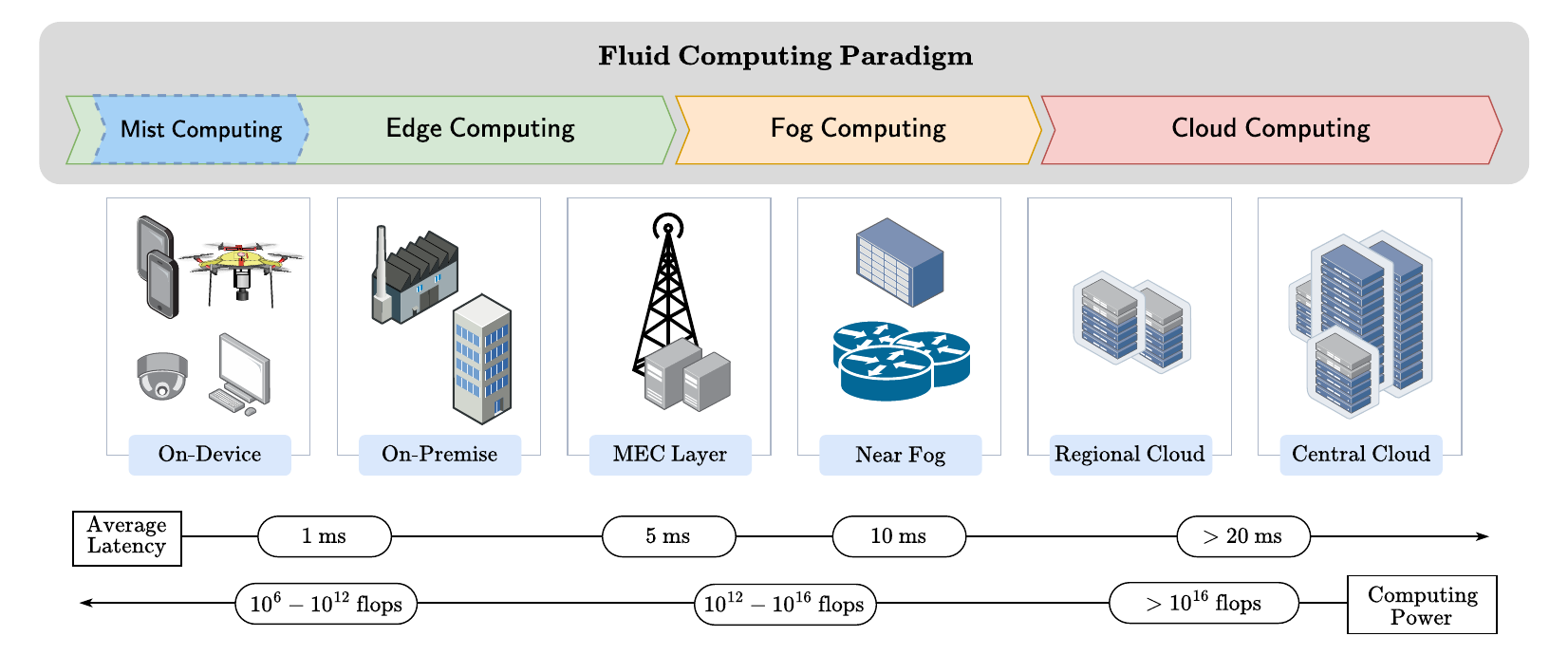}
    \caption{Schematic representation of Fluid Computing paradigm as a unified platform of computing and communication resources that subsumes the main computing paradigms (Mist, Edge, Fog and Cloud). The left-to-right arrow at the bottom indicates increasing delay as data and tasks move toward the cloud, while the right-to-left arrow indicates decreasing computing capabilities as tasks move closer to end-user devices.}
    \label{fig:fluid_computing}
\end{figure*}

Current communication networks consist of a range of heterogeneous computing resources that enable the deployment of distributed applications. These applications often span multiple tiers of computing resources to optimize efficiency and performance, depending on the specific requirements, from end devices and near-edge/fog infrastructure to regional and high-performance cloud resources. In practice, this task-splitting across heterogeneous tiers is often suboptimal and may increase energy consumption and digital carbon footprint due to inefficient communications and data exchanges between tiers~\cite{savazzi2022energy,trihinas2022towards}. Fluid Computing~\cite{al2024computing} (also known as Computing Continuum or Cloud-to-Edge environments) has emerged as a novel distributed computing approach to address these challenges by optimizing application deployment/orchestration across a unified pool of computing and connectivity resources where tasks can transition according to current availability and application requirements (as shown in Figure~\ref{fig:fluid_computing}). 

However, Fluid Computing environments face notable challenges, mostly in terms of orchestration and resource management. Research in this field is scarce, and centralized solutions proposed in the literature are not directly applicable to highly dynamic scenarios~\cite{ullah2023orchestration,gkonis2023survey}. To enable real-world orchestration in these settings, it is essential to address efficient resource allocation and dynamic service deployment based on energy-delay-privacy trade-offs, together with interoperability across administrative domains\footnote{Throughout this paper, we use ``domain'' (or administrative domain) to denote the technical/administrative scope with its own resources and control logic, whereas ``provider'' refers to the organization/business actor that operates one or more such domains (e.g., different regions, business units, or technology domains) that manages contracts, agreements, and governance.} and enabling technologies for lightweight communications and safe workload executions. This interoperability is required because application placement and runtime execution may span multiple tiers and/or administrative domains as conditions change, while still preserving clear authority separation and local autonomy. This viewpoint aligns with the fluid philosophy and abstraction adopted in our prior position paper~\cite{cajaraville2025decentralized}, which treats the environment as a unified platform and relies on multi-domain orchestration and cross-domain runtime execution rather than a single infrastructure layer.

This context motivates architectures where tenants (i.e., application owners/operators) can request high-level deployment intents and domain capabilities to deploy their distributed applications, while domain-side control services can be leveraged to maintain end-to-end (E2E) behavior across domains (e.g., Software-Defined Networking (SDN) policy enforcement and Quality-of-Service (QoS) control). These services create an opportunity for application-level enhancement where the infrastructure can assist applications during runtime execution by exposing controlled services, rather than considering the network as a passive substrate.

However, current multi-domain orchestration proposals still show a weak coupling between orchestration and service programmability: cross-domain runtime execution is often reduced to placement decisions, while domain-side control services are not systematically integrated as a main component for application enhancement~\cite{etsi-zsm-016,hasan2025journey}. Security in Federated Learning (FL) environments is an outstanding example of why domain-side service programmability matters beyond placement. FL is usually deployed across independent organizations to collaboratively train models without sharing raw data, but the training process is vulnerable to model poisoning and Byzantine threats that can degrade the model performance. Recent surveys~\cite{feng2025survey, yuan2024decentralized} emphasize that most common Byzantine-robust mechanisms are evaluated under centralized aggregation assumptions and global visibility. Decentralized Federated Learning (DFL) further complicates this because aggregation is performed through neighbor sharing and local visibility. Proposed DFL defenses~\cite{fang2024byzantine} mainly introduce decentralized robustness mechanisms (e.g., robust aggregation), but typically assume a single administrative domain and do not explore complementary, domain-assisted enforcement mechanisms when the application spans domain boundaries. The question is clear: can domain-side network control services be leveraged as a runtime enforcement surface to enhance DFL security under multi-domain deployments? While existing works propose the intersection of networking control and FL in order to protect networks (e.g., intrusion detection in SDN-enabled systems~\cite{hernandez2025intrusion}), the converse direction is less explored by using SDN programmability to enforce mitigation actions for the DFL learning process under multi-domain settings. There is a lack of approaches that exploit domain-side control services and multi-domain coordination to deliver runtime security enhancement for DFL deployments.

Motivated by the need for multi-domain fluid orchestration and runtime application enhancement, this paper presents the following contributions:

\begin{itemize}
    \item We propose an agnostic multi-domain orchestration architecture for Fluid Computing environments that treats decentralized coordination and domain-side control services as first-class capabilities for intent-driven deployments and fluid runtime execution across the continuum.    
    \item As a representative proof of concept of the proposed architecture, we instantiate it through a concrete multi-domain DFL deployment and define a domain-side workflow for SDN-enabled security enhancement, without introducing a centralized controller or requiring global visibility.
    \item Within this use-case instantiation, we propose FU-HST (Feedback-Updated Half-Space Trees), an SDN-enabled anomaly detection and mitigation algorithm for multi-domain DFL applications that cooperates with Byzantine-robust aggregation techniques to enhance application security at runtime.
    \item We validate the proposed use-case algorithmic solution through DFL simulations over single-domain and multi-domain settings, including Byzantine robustness through detection metrics, DFL performance, and overhead analysis.
\end{itemize}

The remainder of this paper is organized as follows. Section~\ref{sec:related} provides a comprehensive review on multi-domain orchestration and coordination in the computing continuum, and security in (D)FL deployments. Section~\ref{sec:background} outlines our system model and its underlying infrastructure that supports the proposed orchestration architecture. Section~\ref{sec:proposed_framework} elaborates on the proposed multi-domain and decentralized orchestration architecture, including its components and framework workflow. Section~\ref{sec:use-case} introduces the multi-domain DFL system setting and the SDN-enabled anomaly detection mechanism. Section~\ref{sec:experimental_setup} describes the experimental setup and simulation settings, and Section~\ref{sec:results} presents the evaluation results. Finally, Section~\ref{sec:conclusions} concludes the paper, highlighting the contributions and outlining future work.

\section{Related Works}
\label{sec:related}

\subsection{Orchestration in the Cloud–Edge Continuum}

Orchestrating the management and deployment of complex distributed applications is an essential part of heterogeneous computing environments. Frameworks such as Kubernetes, ETSI NFV-MANO (Network Function Virtualization Management and Orchestration), and ONAP (Open Network Automation Platform) are well established in cloud and telecom management stacks. These approaches standardize resource orchestration and service lifecycle, building upon mature architectural blocks.

For example, ETSI NFV-MANO~\cite{etsi2014nfvman001} provides lifecycle management tools and reference points for orchestrating NFVs and cloud-native services, including reference points that allow interoperability among multiple domains. Similarly, ETSI MEC (Multi-access Edge Computing)~\cite{etsi-mec-003} defines an edge platform that divides applications, resource capabilities and management architecture, which extends centralized orchestration in cloud settings. ONAP represents a widely adopted policy-driven orchestration platform for E2E services within operator-centric environments. For example, it has been used in~\cite{rodriguez20205g} to provide network slicing capabilities based on policy-based orchestration solutions to automate virtual (and physical) network functions. Because many of these architectures are designed for centralized and controlled environments, they pose challenges when orchestrating dynamic and distributed scenarios such as edge/mist-located resources and mobile workloads.

More recent research has explored how to orchestrate distributed computing environments on the Cloud-to-Edge axis. For example, \cite{ullah2025towards} proposes a distributed orchestration approach inspired by swarm intelligence principles, allowing the deployment of distributed applications across edge and cloud domains. In~\cite{escobar2023decentralized} is presented an IoT framework for the deployment of serverless applications in Cloud-to-Edge scenarios, based on a decentralized Pub/Sub communication protocol and consisting of a centralized orchestration plane for service deployment and lifecycle management. In~\cite{filinis2024intent}, the authors propose an intent-driven orchestration framework for serverless applications based on autoscaling and scheduling algorithms; however, it considers static edge-cloud tiers. Despite the initiatives, the field of decentralized orchestration (oriented to fluid computing environments) still lacks maturity and standardized solutions, which underscores the need for research on scalable and multi-domain solutions.

In fact, orchestration in multi-domain scenarios must take into account the administrative and policy boundaries among different providers (and their domains). This approach motivates cross-domain coordination mechanisms, which are more effective than globally centralized control mechanisms (e.g., marketplaces). For instance, ETSI ZSM (Zero-touch Network and Service Management)~\cite{etsi-zsm-016} architecture targets E2E network and service management automation through cross-domain interactions, facilitating intent-driven exchanges between domains and closed-loop automation for lifecycle monitoring (not only at deployment time). However, ZSM does not specify application-centric orchestration mechanisms for fluid deployments that jointly coordinate resource placement and application-aware execution, with other domain services (such as SDN or QoS planes) for multi-domain application enhancement.

In summary, current approaches typically focus on either network service management in telecom environments, or workload placement across heterogeneous resources under (usually) a single decision entity. Our work proposes a decentralized architecture that underpins the coordination between per-domain orchestration planes as the main component of a (multi-domain) computing continuum environment. This approach is distinct in that it explicitly links multi-domain coordination to runtime execution requirements, and exposes it across multiple control services (deployment orchestration and network-based services), enabling native architecture-support for application-level enhancement mechanisms.

\subsection{Security in (Decentralized) Federated Learning}

FL paradigm has been identified as susceptible to training-time adversaries through the manipulation of the collaborative/distributed optimization process, which can compromise model performance and system robustness. These threats are often referred to as Byzantine attacks, and it has been demonstrated that malicious model updates can be particularly damaging because they directly target the aggregation mechanism and can be adapted to evade statistical checks. The most prevalent Byzantine attacks can be categorized into two primary groups~\cite{feng2025survey}: data poisoning, which involves manipulating the local data used to train the local model; and model poisoning, which involves manipulating the local model update to skew the overall system performance.

The most studied countermeasures are Byzantine-robust defenses~\cite{shi2022challenges}, which employ different techniques (e.g., robust statistics, similarity-based filtering, or reputation-driven weighting) to strengthen the aggregation process and mitigate or erase malicious updates. However, most robust aggregation mechanisms assume a coordination by a unique point of control in centralized FL settings and may not be as effective under strong non-IID conditions. On the other hand, in DFL settings, each client aggregates only the updates received from its neighbors. This results in a weaker form of Byzantine robustness, characterized by local bias and topology-dependency. Recent proposals explicitly address this setting by designing robust neighbor aggregation rules suitable for sparse networks. For instance, BALANCE~\cite{fang2024byzantine} provides a robust decentralized learning rule that performs similarity-based neighbor filtering to enable Byzantine-robust local averaging with convergence guarantees.

In addition to aggregation-based defenses, several proposals incorporate alternative mechanisms to improve attacker identification. For instance, SIREN~\cite{guo2024siren} introduces proactive alarming procedures that allow the FL server to identify which clients suffer global model deviation and to identify where abnormal contributions could come from. Other approaches involve exploring infrastructure-assisted FL, in which network programmability is considered to improve learning performance. For example, the work~\cite{mahmod2023improving} leverages the SDN paradigm to enhance the FL application by adapting networking services rather than relying exclusively on learning-side changes. However, recent DFL studies~\cite{feng2024dart} have highlighted that mechanisms designed for centralized FL may not transfer cleanly to DFL. Therefore, Byzantine robustness must be analyzed taking into account the decentralized nature of DFL settings and networking considerations arising from peer-to-peer communications and dynamic topologies.

In summary, Byzantine-robust defenses are evaluated primarily in centralized FL settings, and common assumptions (global visibility) do not directly carry over to decentralized topologies. Even under DFL settings, recent proposals~\cite{fang2024byzantine,cajaraville2025byzantine} typically model a single administrative domain, i.e., they do not consider coordination when neighbor graph spans multiples domains. Furthermore, SDN-assisted proposals place a strong emphasis on FL quality enhancement within single-domain networking actions, rather than domain-side mechanisms for runtime enforcement across multiple administrative domains. Our work addresses these gaps by proposing an SDN-enabled anomaly detection algorithmic solution tailored to multi-domain DFL settings. Specifically, each domain runs an SDN application that scores participants from local-generated alerts and coordinates with different SDN domains to keep Byzantine mitigation consistent E2E, without requiring global visibility.

\section{System Model}
\label{sec:background}

Our aim is to transform the heterogeneous continuum of resources into a reliable and unified computing fabric operated across multiple administrative domains. The system model under consideration involves: (i) tenants that request the deployment and execution of distributed applications; (ii) applications composed of components and associated workflows, together with their functional and non-functional requirements; (iii) heterogeneous computing and communication resources distributed across multiple administrative domains; and (iv) domain-local control entities that observe resource status and drive deployment and runtime decisions. Together, these elements define the operational environment of the Fluid Computing paradigm: tenants request the execution of applications over resources owned by different domains, where workloads run under domain authority, while domain-side control entities decide how such applications are deployed, adapted, and maintained across domains over time.

At an abstract level, we model a fluid computing environment as a multi-domain system composed of three interacting planes that provide complementary views of the same environment: (i) a data plane for data-centric communication; (ii) a virtualization plane for safe multi-tenant execution; and (iii) an orchestration plane that observes, decides, and actuates over distributed resources managed by different domains. The first two planes provide the baseline substrate for data exchange and workload execution, whereas the orchestration plane provides the control logic that coordinates how such resources and execution contexts are used to satisfy application requirements under domain constraints.

\textbf{Data Plane.}
This plane models the communication fabric through which application components exchange data, control information, and event notifications across domains. In particular, modern IoT and AI workflows are centered on time-sensitive data streams rather than request-response transactions. Publish/Subscribe (Pub/Sub) protocols, therefore, dominate such applications because they decouple producers from consumers in space, time, and synchronization. In our model, we assume a distributed Pub/Sub substrate that supports data communication in terms of data keys rather than host address (aligned with concept Named Data Networking~\cite{zhang2014named}), which makes it suitable for fluid computing environments, where publishers, subscribers, and even the topology itself may migrate continuously. Solutions such as Zenoh~\cite{zenoh2021} exemplify this design space.

\textbf{Virtualization Plane.}
This plane models the execution substrate that hosts application components on concrete resources under domain authority. In particular, fluid environments require safe multi-tenant execution across heterogeneous resources and continuous visibility of resource availability. In our model, nodes expose resources telemetry to enable orchestration decisions and isolate workloads from the host operating system and (potentially) mutually untrusted tenants. Lightweight virtualization technologies such as WebAssembly~\cite{rossberg2021webassembly} exemplify this approach by enabling sandboxed execution with a policy-controlled system-call surface. We also assume that nodes run a middleware that manages local execution and interfaces with the orchestration plane.

\textbf{Orchestration Plane.}
This plane models the control logic that governs how applications are deployed, adapted, and maintained over time across domain-owned resources by translating intent-based requests from tenants into deployment and runtime actions over the available resources. In our model, orchestration is inherently multi-domain since each domain manages its own administrative boundary and exposes controlled interfaces for coordination with peer domains for jointly optimizing placement decisions and resource utilization across the continuum to meet application requirements. This philosophy aligns with E2E cross-domain service management concepts in ETSI ZSM~\cite{etsi-zsm-016} and with the multi-domain interoperability exchanges defined in ETSI NFV-MANO~\cite{etsi2014nfvman001}.

In this paper, we focus on our vision of the orchestration plane and its multi-domain coordination capabilities (discussed in Section~\ref{sec:proposed_framework}); therefore, Data Plane and Virtualization Plane are abstracted as the enabling substrate infrastructure for the higher-level orchestration mechanisms.

\section{Agnostic Decentralized Orchestration Architecture for Fluid Computing}
\label{sec:proposed_framework}

This section introduces an orchestration architecture for deploying and executing distributed applications (e.g., AI or IoT applications) across multi-domain fluid computing environments, i.e., creating a unified platform/fabric of resources. The architecture is agnostic to the specific application being deployed, and defines the main architectural elements and their roles (in Subsection~\ref{subsec:architecture}), 
the E2E workflow from application request to runtime execution (in Subsection~\ref{subsec:workflow}), and the orchestration-plane details that enable decentralized coordination among different domains and application-level enhancement via domain-side control services (in Subsection~\ref{subsec:coordination}).

\subsection{Architecture Overview}
\label{subsec:architecture}

\begin{figure*}[tbp]
    \centering
    \includegraphics[width=1.0\linewidth]{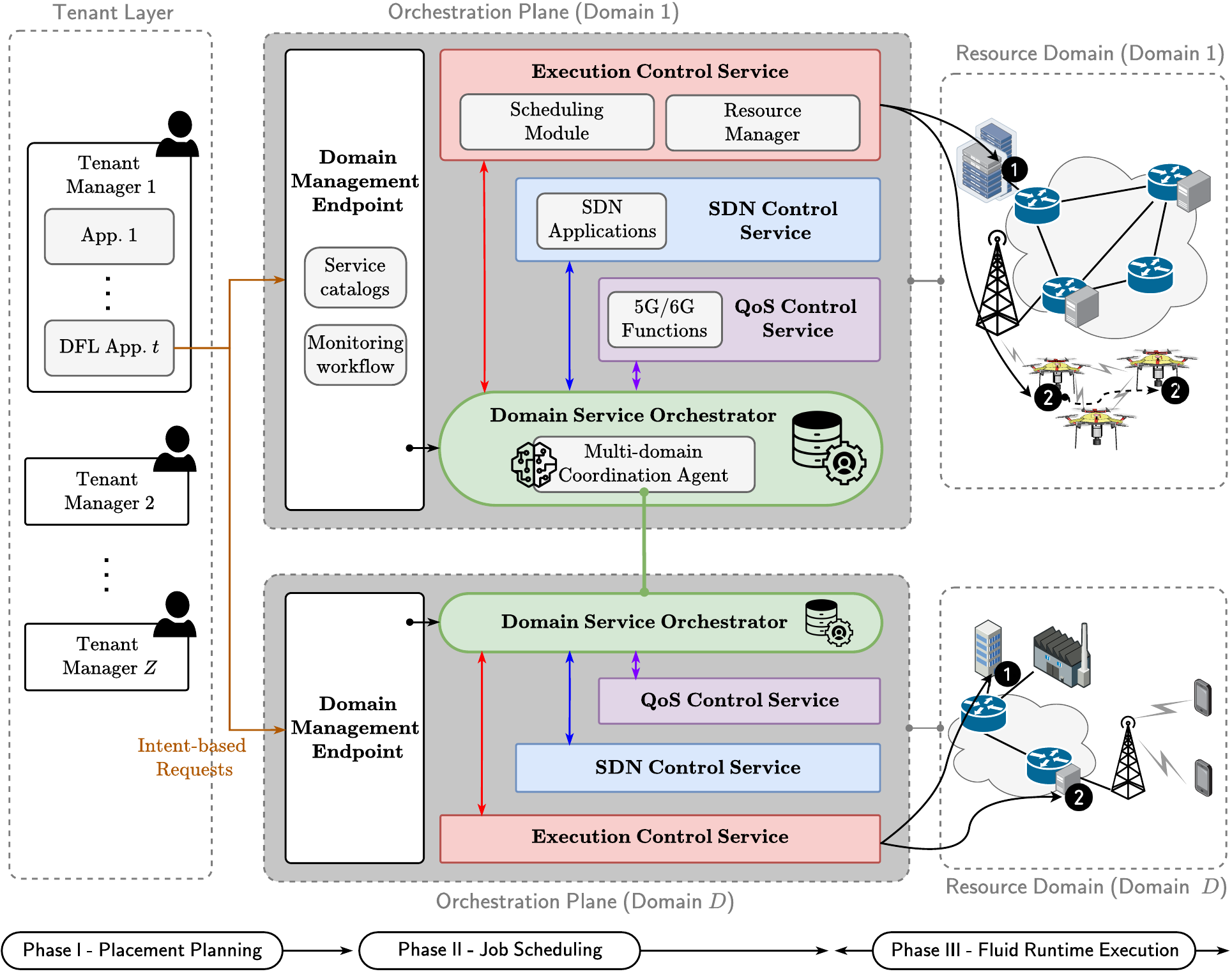}
    \caption{Agnostic multi-domain orchestration architecture for multi-tenant Fluid Computing environments, enabling decentralized coordination across administrative domains.}
    \label{fig:fluid_architecture}
\end{figure*}

Figure~\ref{fig:fluid_architecture} depicts the proposed architecture as a unified fabric organized into three parts: a multi-domain Orchestration Plane as the central element of the architecture, a Tenant Layer (on the left side), and per-domain Resource Domains (on the right side). The main objective of this separation is to let tenants decide about what they need (intents and capabilities through different interfaces) to deploy their application, while the orchestration plane (in fact, the domains) decides how to realize it within local policy boundaries. This philosophy is aligned with the intent-based networking concepts and definitions~\cite{rfc9315}.

In the \textbf{Orchestration Plane}, each domain consists primarily of a Domain Service Orchestrator (\textsc{dso}), which coordinates the domain-side control logic and decides how tenant intents are translated into actions through its domain-side control services, including the Execution, SDN and QoS control services. When required, a Multi-Domain Coordination Agent (\textsc{mdca}), associated with the \textsc{dso}, coordinates with peer domains under controlled interfaces to fulfill cross-domain actions. Furthermore, each domain exposes a Domain Management Endpoint component (\textsc{dme}) that is capability-driven and does not disclose internal information about the own Resource Domain, acting as the main northbound interface between tenants and the Orchestration Plane. It offers resource capability abstraction (computational and memory/storage availability, regions...), service catalogs (security and networking policies, QoS offerings), lifecycle management and policy boundaries. It also validates admission constraints of tenants.

On the left side, organized as the \textbf{Tenant Layer}, tenants model each distributed application as a set of deployable components/containers/nanoservices with dependencies and requirements. These descriptors capture relevant network- and runtime-level intents, including hardware resource demands, elastic QoS envelopes (e.g., deadlines), trust and privacy constraints (e.g., requiring workloads or data to remain within specific domains), runtime policies (e.g., whether migration is permitted), or security posture (e.g., enabling runtime isolation via domain SDN services). Each tenant operates a Tenant Manager component as the logical control component responsible for interacting with domains and assembling deployment requests.

On the right side, each \textbf{Resource Domain} spans heterogeneous assets (IoT devices, vehicular platforms, on-premise capabilities, fog nodes and/or cloud regions), and the nature of each domain may differ among them. For example, some domains are cloud-only, other are edge-heavy radio-access domains, and other are enterprise/private infrastructures. The key idea is that infrastructure diversity is abstracted through the \textsc{dme} and the Orchestration Plane, enabling tenants to reason at the level of capabilities and intents rather than infrastructure specifics. Furthermore, the Data Plane and Virtualization Plane are assumed within the Resource Domain as the enabling substrate for safe execution and telemetry (as modeled in Section~\ref{sec:background}).

\subsection{Framework Workflow}
\label{subsec:workflow}

We establish the E2E deployment and execution workflow into three phases as previously shown in Figure~\ref{fig:fluid_architecture}. The phases indicate the role of each component from the tenant request to runtime execution.

\subsubsection{Phase I -- Placement Planning}
Firstly, the tenant submits the application descriptor and its intents/constraints to the Tenant Manager. In order to take proper domain selections and intent decompositions, the Tenant Manager continuously gathers data from domains with which it has agreements. This data, which includes domain catalog information and high-level resource availability, is offered by the corresponding \textsc{dme}s and is employed to satisfy business and policy constraints (e.g., contracts or cost). Possible driver decisions include geographic proximity (latency-aware tasks near users), data locality, resilience independence (avoid a single domain dependency), and heterogeneous capabilities (e.g., one domain offers dense MEC presence but another offers cloud accelerators). The output of the Tenant Manager is a multi-domain placement intent, specifying which components are requested in each selected domain and the required domain-side service capabilities (e.g., QoS/slice and SDN services).

\subsubsection{Phase II -- Job Scheduling}
Each domain independently schedules the components assigned to its domain through its orchestration plane (\textsc{dso} and control services). This includes mapping application components to resource pools and provisioning the requested connectivity capabilities and service endpoints. At this point, inter-domain coordination is limited to establishing cross-domain continuity (e.g., bindings and reachability for cross-domain dependencies) via the \textsc{mdca}s.

\subsubsection{Phase III -- Fluid Runtime Execution}
This phase captures the core ``fluid'' philosophy within this architecture, i.e., adapting application execution at runtime across heterogeneous resources and domains under mobility, workload and resource variation. When a domain-local runtime event threatens continuity (or enables a better feasible execution elsewhere), the \textsc{dso} triggers its \textsc{mdca} to negotiate the required cross-domain action with peer domains. Specifically, this phase considers mechanisms such as component offloading/migration (within its own domain or across peer domains), elastic scaling, and network/policy reconfiguration so that the E2E service continues to satisfy its intent under dynamic conditions. For instance, fluid runtime execution can be exploited in 6G-based settings where nearby devices form self-organized D2D clusters, enabling opportunistic offloading to transient local compute pools as conditions vary.

\subsection{Decentralized and Multi-Domain Orchestration Plane}
\label{subsec:coordination}

This subsection details the internal control services coordinated by the \textsc{dso} and the behavior of the \textsc{mdca}s introduced in Figure~\ref{fig:fluid_architecture}. Within each domain, the \textsc{dso} receives intent-based requests from the \textsc{dme} and translates them into domain-governed actions by delegating to domain-side control services and supervising their execution. The \textsc{dso} is also the main point for decentralized coordination through the \textsc{mdca}, which exchanges constrained coordination messages with peer domains (specifically, other \textsc{mdca}s) to preserve cross-domain continuity and runtime execution enhancement. This coordination is not treated as a pre-scheduling step; instead, it is a runtime capability that propagates local decisions beyond the administrative boundary. The decentralized nature is considered since no single provider has global authority over resources and policies, and a centralized marketplace introduces scalability bottlenecks and governance/trust challenges. Thus, this multi-domain orchestration plane (with its control services) is foundational since it provides explicit orchestration and coordination capabilities at the architectural level to satisfy workload execution, traffic configuration, and runtime continuity.

\textbf{Execution Control Service:} it realizes domain-local control of application execution on domain-owned resources. It is composed of a Scheduling Module for assigning workloads to local resources, and a Resource Manager for maintaining abstracted resource state and execution domains (through the Virtualization Plane). It consumes translated intents from the \textsc{dso}, validates policy constraints (e.g., data locality, trusted execution domains), and schedules application components to resources. When the current domain cannot maintain the intent locally (e.g., cross-domain migration triggered by mobility, and job relaying when policy constraints prevent local admission), it exposes execution-level hooks that allow the \textsc{dso}, via its \textsc{mdca}, to initiate cross-domain relocation. In this case, the coordination agent negotiates with peer domains by issuing a relocation request carrying an abstract capability and constraint profile (including admissible policies and SLA bounds). Then, peer coordination agents respond with feasibility decisions and, when accepted, a deployment handle that enables the requesting \textsc{dso} to complete the cross-domain action while updating component bindings and dependencies consistently.

\textbf{SDN Control Service:} it provides programmability of forwarding and policy enforcement (e.g., traffic filtering and segmentation, overlay establishment) within the domain boundaries. It executes SDN applications that can be domain-driven (e.g., enforcing domain policies and meeting SLAs) or tenant-requested under constrained interfaces, where high-level intents are translated into enforceable domain policies. Under multi-domain deployments, SDN Control Service supports cross-domain reachability and connectivity consistency when components communicate across domains (e.g., overlay stitching and border policy alignment), coordinated by the \textsc{dso} through multi-domain coordination so that runtime execution preserves the application intent.

\textbf{QoS Control Service:} it denotes the set of services that provide QoS assurance and network slicing/connectivity guarantees, typically realized through 5G/6G management and orchestration capabilities. Exposing QoS/slice capabilities as lifecycle-managed services is aligned with slicing management frameworks that describe preparation, commissioning, operation, and decommissioning phases for network slice instances. This service may be consumed by domain actions (e.g., mobility-aware continuity) or tenant-based intents (e.g., request a slice template for a specific application), subject to domain policies. When an application spans domains, QoS Control Service preserves continuity of traffic treatment across borders (e.g., QoS/slice mapping and admission alignment), coordinated through the \textsc{mdca} and peer domains to maintain E2E behavior under runtime variability.

\section{Use Case of Security Enhancement in Multi-Domain Decentralized Federated Learning}
\label{sec:use-case}

This section instantiates the proposed architecture of Section~\ref{sec:proposed_framework} through an operational use case based on a multi-domain DFL application deployment, and uses this setting to demonstrate how domain-side control services can support runtime application enhancement. Concretely, we focus on a domain-side security mechanism that operates during execution to mitigate both intra- and cross-domain Byzantine threats. The remainder of this section maps the general architecture into a simplified system model tailored to this specific deployment (Subsection~\ref{subsec:use_case}), defines the DFL system setting and proposed runtime workflow (Subsection~\ref{subsec:system_setting}), and describes the domain-side anomaly detection methodology (Subsection~\ref{subsec:anomaly_detection}).

\subsection{From Proposed Architecture to Use Case}
\label{subsec:use_case}

In this section we consider a use-case-based simplification of the proposed architecture that maintains its overall philosophy, while abstracting several mechanisms to focus on the effectiveness of fluid runtime enhancement via domain-side control services and decentralized coordination. We adopt a DFL workflow, a paradigm that naturally aligns with multi-domain deployments since participants can be distributed across heterogeneous administrative domains, while the learning process explicitly relies on peer-to-peer model exchange over a communication graph that may include cross-domain neighbor relationships.

The overall architecture remains unchanged: tenants consume domain capabilities through management interfaces, the application spans multiple administrative domains, and each domain operates its own orchestration plane to manage local workloads on its Resource Domain. However, our methodological focus is the SDN Control Service as a concrete domain capability used for runtime enhancement. SDN provides connectivity and policy primitives that are directly relevant to multi-domain execution, while allowing domains to preserve local autonomy and confidentiality. Specifically, the tenant requests a security-enhanced SDN application that gathers information and alerts produced by participants during decentralized training in order to generate filtering directives that mitigate Byzantine attacks from malicious participants. Notably, these mitigation actions can be enforced under multi-domain governance constraints. 

To keep the evaluation centered on the application-level logic, we abstract mechanisms that would otherwise dominate the methodology (e.g., tenant placement optimization, agent-logic for cross-domain coordination, and slice instantiation workflows). Instead, we assume the multi-domain DFL deployment is instantiated, and we focus on runtime execution, both the decentralized learning process and the SDN-enabled security mechanism.

\subsection{System Setting}
\label{subsec:system_setting}

\begin{figure*}[tbp]
    \centering
    \includegraphics[width=0.8\linewidth]{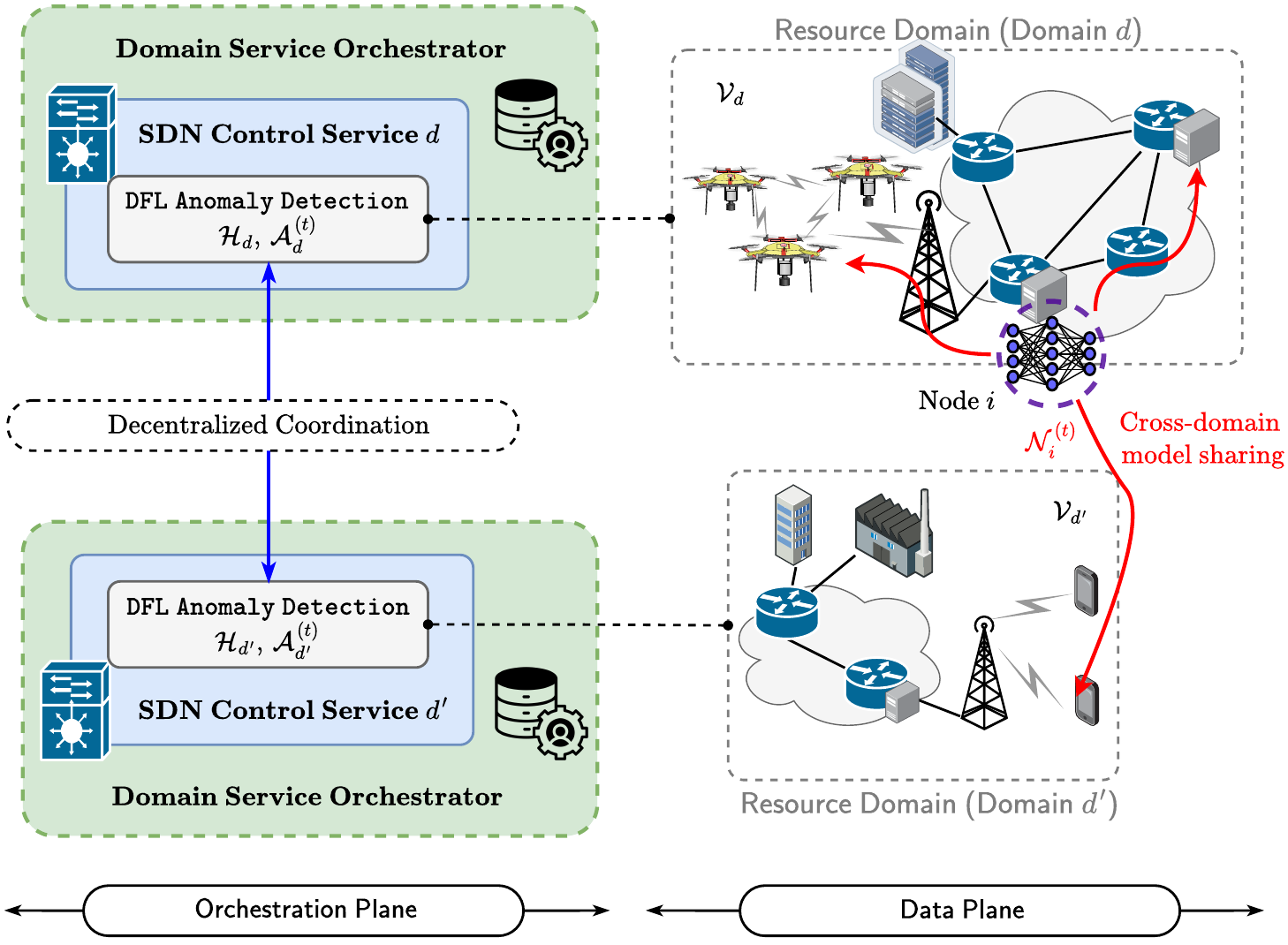}
    \caption{Considered scenario for multi-domain DFL deployment with SDN-enabled security-enhancement mechanism}
    \label{fig:system_setting}
\end{figure*}

We consider a DFL system (which is shown in Figure~\ref{fig:system_setting}) with a set of clients $\mathcal{V} = \{1,\dots,N\}$ deployed across $D$ administrative domains, i.e., each domain $d \in \{1,\dots,D\}$ owns a subset $\mathcal{V}_d$ such that $\mathcal{V} = \bigcup_{d=1}^{D} \mathcal{V}_d$ and $\mathcal{V}_d \cap \mathcal{V}_{d'} = \emptyset$ for $d \neq d'$. Clients perform a distributed collaborative training of a Machine Learning (ML) model following the DFL paradigm, where each node $i \in \mathcal{V}$ exchanges model updates only with a neighbor subset $\mathcal{N}_i^{(t)} \subseteq \mathcal{V} \setminus \{i\}$. This neighborhood relationship is defined by a communication graph $G=(\mathcal{V},\mathcal{E})$ that can span multiple domains, where $\mathcal{E} \subseteq \mathcal{V} \times \mathcal{V}$ represents the set of communication links between clients (including intra- and inter-domain neighbors).

\textbf{Per-round DFL workflow:} DFL workflow is carried out over synchronous rounds $t = 1,\dots,T$. At each round $t$, every node $i$ performs local training on its private data to update its local model (training step). Then, it exchanges the model parameters update with its neighbors (sharing step). Finally, node $i$ aggregates the set of received updates together with its local model using a pre-defined aggregation rule to obtain its next-round local model.

\textbf{Threat model:} We assume a subset $\mathcal{M} \subset \mathcal{V}$ of malicious nodes that can behave arbitrarily in the DFL protocol, attempting to degrade convergence or bias the collaboratively learned model. We assume that malicious nodes follow rigorously the previous DFL workflow (i.e., train the local model, share update with their neighbors and aggregate received updates following the aggregation rule) but they can manipulate the shared model update (e.g., through model/data poisoning updates).

\textbf{Alert generation:} we assume that the aggregation rule enables nodes to identify which neighbor updates are trustworthy or potentially malicious/biased. This idea is related to Byzantine-robust aggregation rules that employ filtering techniques, or proposals like~\cite{cajaraville2025byzantine} that use different strategies to weigh the trustworthiness of received updates. Therefore, at round $t$, node $i$ produces a lightweight alert signal, denoted by $w_{ij}^{(t)} \in [0,1]$, about each received update from neighbor $j \in \mathcal{N}_i^{(t)}$. This signal represents node $i$'s assessment of model update from neighbor $j$ at round $t$. Thus, each node produces an alert vector $( w_{ij}^{(t)})_{j \in \mathcal{N}_i^{(t)}}$ once per round after the aggregation step.

\textbf{SDN-enabled control logic:} Each domain $d$ runs an SDN application called \texttt{DFL Anomaly Detection}, attached to the corresponding SDN Control Service, that receives the generated alerts of nodes from $\mathcal{V}_d$. Since alerts may originate from both intra- or inter-domain neighbors, per-domain SDN applications coordinate among them to relay inter-domain alerts in order to reconstruct the overall set of alerts received for each node $i \in \mathcal{V}_d$. Then, each domain independently processes the alerts and outputs a round-based estimated anomalous set $\mathcal{A}_d^{(t)} \subseteq \mathcal{V}_d$, which is employed to return a mitigation action to clients or network (in our case, a ban list). Upon receiving the SDN response at round $t$, in subsequent round $t+1$, node $i$ filters any model update originating from a banned node before running its aggregation procedure. Although the ban list is recomputed at every round from the current alert stream, the SDN-enabled control logic incorporates temporal stabilization and feedback mechanisms that facilitate the repeated identification of persistently malicious nodes across subsequent rounds.

\subsection{SDN-enabled Anomaly Detection Algorithm}
\label{subsec:anomaly_detection}

\begin{figure}[tbp]
    \centering
    \includegraphics[width=0.43\linewidth]{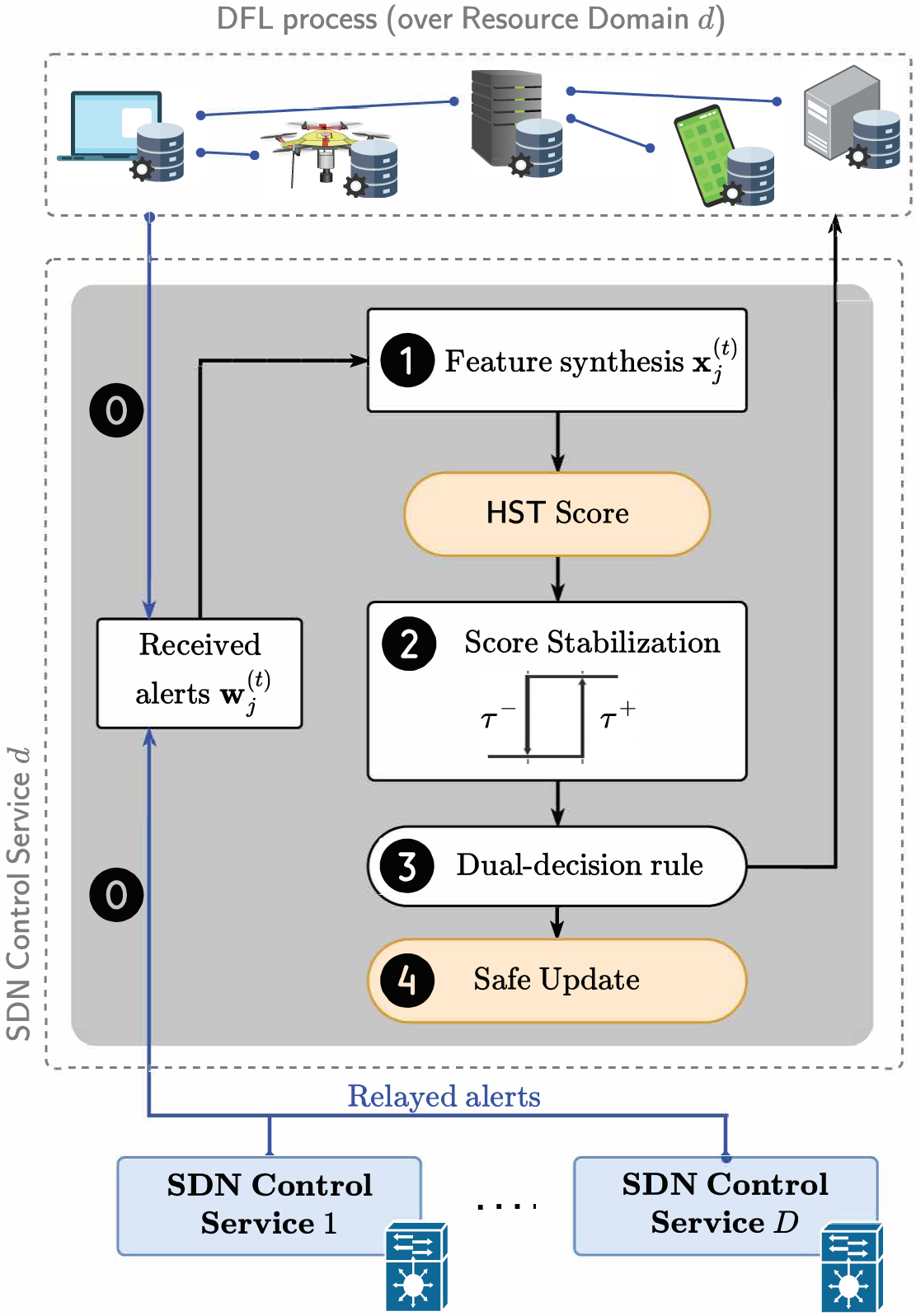}
    \caption{Overview of the SDN-enabled anomaly detection workflow (FU-HST algorithm)}
    \label{fig:SDN_anomaly_detection}
\end{figure}

We model anomaly detection mechanisms as a round-based streaming problem in which the SDN application ingests the newly generated alerts, produces an online anomaly score for each node, and classifies nodes to drive the mitigation action. We propose FU-HST (Feedback-Updated Half-Space Trees), an SDN-enabled anomaly detection algorithmic solution (which is described in Algorithm~\ref{alg:fu_hst} and summarized in Figure~\ref{fig:SDN_anomaly_detection}) that operates once per DFL-round and performs streaming anomaly scoring to identify clients whose behavior deviates persistently from the baseline of the federation. This solution relies on Half-Space Trees algorithm~\cite{tan2011fast}, which is designed for evolving data streams and can operate in an online fashion with bounded per-sample cost.

\begin{algorithm}[tbp]
\LinesNumbered
\caption{FU-HST (round $t$, domain $d$)}
\label{alg:fu_hst}
\KwIn{
model $\mathcal{H}_d$, input alerts $\{w_{ij}^{(t)}\}$, previous state $(\mathbf{f}^{(t-1)}, \mathbf{s}^{(t-1)}, \mathbf{c}^{(t-1)})$ of domain $d$, hyperparameters $\tau, \gamma, \alpha, \beta \in (0,1)$, $p_\text{u} \in [0,1]$, 
}
\KwOut{
$\{y_j^{(t)}\}_{j \in \mathcal{V}_d}$, $\mathcal{A}_d^{(t)}$, updated model $\mathcal{H}_d$, updated state $(\mathbf{f}^{(t)}, \mathbf{s}^{(t)}, \mathbf{c}^{(t)})$.
}

\BlankLine
$\mathcal{A}_d^{(t)} \leftarrow \emptyset$\\
$\tau^{+} \leftarrow \tau$\\ 
$\tau^{-} \leftarrow \gamma \tau$\\

\ForEach{$j\in\mathcal{V}_d$}{
    \tcp{Feature synthesis}
    $\mathcal{R}_j^{(t)} \leftarrow \{ i \in \mathcal{V}: j \in \mathcal{N}_i^{(t)} \}$\\
    $\mathbf{w}_j^{(t)} \leftarrow (w_{ij}^{(t)})_{ i \in \mathcal{R}_j^{(t)}}$\\

    Compute $\overline{w}_j^{(t)}$ and $\overline{z}_j^{(t)}$ as Eq. \ref{eq:mean_alert} and \ref{eq:zscore_alert}, respectively\\
    
    $\mathbf{x}_j^{(t)} \leftarrow (\mathbf{w}_j^{(t)}, \overline{w}_j^{(t)}, \overline{z}_j^{(t)}, f_j^{(t-1)})$

    $\tilde{y}_j^{(t)} \leftarrow \mathsf{Score} \left(\mathcal{H}_d, \mathbf{x}_j^{(t)} \right)$

    \tcp{Score stabilization}
    \uIf{$\tilde{y}_j^{(t)} < \tau^{-}$}{
        $(s_j^{(t)}, f_j^{(t)}, c_j^{(t)}) \leftarrow (0,0,0)$
    }\Else{
        $s_j^{(t)} \leftarrow \alpha s_j^{(t-1)}+(1-\alpha) \tilde{y}_j^{(t)}$\\
        $f_j^{(t)} \leftarrow f_j^{(t-1)}$\\
        $c_j^{(t)} \leftarrow c_j^{(t-1)} + \mathbf{1} \left(\tilde{y}_j^{(t)} > \tau^{+} \right)$
    }

    $y_j^{(t)} \leftarrow s_j^{(t)}$

    \tcp{Dual-decision rule}
    \If{$(y_j^{(t)} > \tau^{+}) \lor (\tilde{y}_j^{(t)} > \tau^{+} \land f_j^{(t-1)} > \tau^{+})$}{
        $\mathcal{A}_d^{(t)} \leftarrow \mathcal{A}_d^{(t)} \cup \{j\}$\\
        $f_j^{(t)} \leftarrow \beta \tilde{y}_j^{(t)} + (1-\beta) f_j^{(t-1)}$
    }
    \tcp{Safe update}
    \uIf{$(y_j^{(t)}=0) \land (c_j^{(t)}=0)$}{
        $\mathcal{H}_d \leftarrow \mathsf{Train} \left(\mathcal{H}_d, \mathbf{x}_j^{(t)}\right)$
    }\ElseIf{$y_j^{(t)} < \tau^{+}$}{
        \If{$U \sim \mathsf{Unif}(0,1) \le p_\text{u}$}{
            $\mathcal{H}_d \leftarrow \mathsf{Train} \left(\mathcal{H}_d, \mathbf{x}_j^{(t)} \right)$
        }
    }
}
\Return{$\{y_j^{(t)}\}_{j \in \mathcal{V}_d},\ \mathcal{A}_d^{(t)}$}
\end{algorithm}

At round $t$, the $d$-domain SDN application gathers both local alerts sent by nodes in $\mathcal{V}_d$,
\begin{equation}
    \mathcal{L}_d^{(t)} = \{ w_{ij}^{(t)} : i \in \mathcal{V}_d, \ j \in \mathcal{N}_i^{(t)} \},
\end{equation}
and relayed alerts sent by nodes outside $\mathcal{V}_d$ toward nodes in $\mathcal{V}_d$, 
\begin{equation}
    \mathcal{I}_d^{(t)} = \{ w_{ij}^{(t)} : i\notin\mathcal{V}_d,\ j\in\mathcal{V}_d,\ j\in\mathcal{N}_i^{(t)}\}.
\end{equation}
This information enables the SDN application to construct, for each $j\in\mathcal{V}_d$, the vector of received alerts
\begin{equation}
    \mathbf{w}_j^{(t)} = (w_{ij}^{(t)})_{i \in \mathcal{R}_j^{(t)}},
\end{equation}
where 
\begin{equation}
    \mathcal{R}_j^{(t)} = \{ i \in \mathcal{V} : j \in \mathcal{N}_i^{(t)}\}
\end{equation}
is the set of nodes that rated node $j$ at round $t$. 

The proposed SDN-enabled anomaly detection loop decomposes into three steps (for each $j \in \mathcal{V}_d$):\\

\textbf{Step I -- Feature synthesis:} From received alerts $\mathbf{w}_j^{(t)}$, the SDN application computes statistics such as the mean received score
\begin{equation}
\label{eq:mean_alert}
    \overline{w}_j^{(t)} = \frac{1}{|\mathcal{R}_j^{(t)}|} \sum_{i \in \mathcal{R}_j^{(t)}}  w_{ij}^{(t)}
\end{equation}
for capturing the consensus of neighbors' assessments of $j$, and the mean standardized deviation 
\begin{equation}
\label{eq:zscore_alert}
    \overline{z}_j^{(t)} = \frac{1}{|\mathcal{R}_j^{(t)}|} \sum_{i \in \mathcal{R}_j^{(t)}} \frac{ w_{ij}^{(t)} - \overline{w}_j^{(t)} }{\sigma_j^{(t)}+\varepsilon},
\end{equation}
where $\varepsilon >0$ and 
\begin{equation}
    \sigma_j^{(t)} = \sqrt{\frac{1}{|\mathcal{R}_j^{(t)}|} \sum_{i \in \mathcal{R}_j^{(t)}}(w_{ij}^{(t)} - \overline{w}_j^{(t)})^2}.
\end{equation}
for capturing the dispersion via $Z$-score normalization.
Then, it assembles the following feature vector for improving robustness under heterogeneous rates and per-round variability:
\begin{equation}
    \mathbf{x}_j^{(t)} = (\mathbf{w}_j^{(t)}, \overline{w}_j^{(t)}, \overline{z}_j^{(t)}, f_j^{(t-1)}),
\end{equation}
where $f_j^{(t-1)}$ is a historical feedback term.\\

\textbf{Step II -- Score stabilization:} The anomaly score is computed by the streaming HST model $\mathcal{H}_d$, i.e.,
\begin{equation}
    \tilde{y}_j^{(t)} = \mathsf{Score}(\mathcal{H}_d, \mathbf{x}_j^{(t)}),
\end{equation}
using a windowed update mechanism to track evolving behavior in non-stationary environments. To handle per-round variability and instability due to isolated spikes, we apply a temporal stabilization step to the raw HST score $\tilde{y}_j^{(t)}$. This mechanism is based on the hysteresis phenomenon (equivalent to the Schmitt-trigger in electronics) with two switching thresholds ---an upper threshold $\tau^{+}$ to assert anomaly and a lower threshold $\tau^{-} = \gamma\tau^{+}$, with $\gamma\in(0,1)$, to clear it---. Thus, it introduces a stable intermediate region that prevents chattering under noisy inputs and supports gradual restoration of trust once evidence of anomalous behavior disappears. We maintain a smoothed score buffer $s_j^{(t)}$ using an exponential moving average (EMA),
\begin{equation}
    s_j^{(t)} = \alpha s_j^{(t-1)} + (1-\alpha)\tilde{y}_j^{(t)},
\end{equation}
for $\alpha \in (0,1)$, and update a maliciousness counter as
\begin{equation}
    c_j^{(t)} = c_j^{(t-1)} + \mathbf{1}\left( \tilde{y}_j^{(t)} > \tau^{+} \right)
\end{equation}
whenever the raw score is not in the highly trustworthy regime, i.e., $\tilde{y}_j^{(t)} \ge \tau^{-}$. Otherwise (if $\tilde{y}_j^{(t)} < \tau^{-}$), we reset the state variables and feedback to $(s_j^{(t)}, f_j^{(t)}, c_j^{(t)}) \leftarrow (0,0,0)$. This reset provides the recovery mechanism in which, if anomalous evidence disappears, the node state is cleared and it is naturally omitted from future ban lists, which supports recovery from false positives.\\

\textbf{Step III -- Dual-decision rule:} Node $j$ is added to $\mathcal{A}_d^{(t)}$ if: (i) the stabilized score $s_j^{(t)}$ crosses the upper threshold; or (ii) the raw score crosses $\tau^{+}$ while the feedback state already indicates prior maliciousness, $f_j^{(t-1)} > \tau^{+}$. Specifically, when a node is flagged, the feedback state is updated as an EMA,
\begin{equation}
    f_j^{(t)} = \beta \tilde{y}_j^{(t)} + (1-\beta) f_j^{(t-1)},
\end{equation}
for $\beta \in (0,1)$, so that repeated evidence is remembered across rounds. Note that the anomalous set $\mathcal{A}_d^{(t)}$ is recomputed at every round. Therefore, persistence occurs only if node $j$ is flagged again in subsequent rounds, which is facilitated by the stabilized score $s_j^{(t)}$ and feedback state $f_j^{(t)}$.

\textbf{Step IV -- Safe update:} Finally, the model $\mathcal{H}_d$ is updated only for nodes classified as strictly safe, which regularizes drift and preserves sensitivity to emerging anomalies. Additionally, to reduce catastrophic forgetting, we admit a small fraction of low-risk alerts into the update stream with probability $p_\text{u}$.

\section{Experimental Setup}
\label{sec:experimental_setup}

This section describes the experimental setup used for the validation and analysis tests. It includes technical details related to ML models, datasets, and algorithm parameters, as well as threat models and algorithmic baselines with which to compare our proposal. Additionally, we introduce the validation metrics used to assess the performance of the anomaly detection algorithm and DFL under different scenarios. 

\subsection{Validation System}
\label{subsec:validation-system}

\textbf{Communication Model:} Communication between clients is modeled as a $D$-block Stochastic Block Model (\textsc{sbm}): (i) intra-domain Erd\H{o}s-R\'enyi graphs, in which each pair of clients from the same domain are neighbors with probability $p_1$, and (ii) an inter-domain Erd\H{o}s-R\'enyi graph, in which each pair of clients from different domains are neighbors with probability $p_2$. This setup makes it easy to observe how interconnections between nodes affect the performance of the system.

\textbf{Learning task:} The nodes perform state-of-the-art training tasks based on image classification using the well-known MNIST dataset~\cite{LeCun10}. This dataset is independently and identically distributed (IID) across all clients. The clients train on the dataset using a convolutional neural network (CNN) consisting of seven layers, including convolutional, pooling, and fully connected layers. This configuration allows us to focus on the main point of this experimental setup, which is the effectiveness of the SDN-enabled anomaly detection algorithm, rather than delving into the learning aspects that are outside the scope of this study. Training is performed over $20$ synchronous rounds. In each round, the clients train their local model over the entire local training set with a learning rate of $0.01$.

\textbf{Byzantine-robust aggregation rule:} We use the Byzantine-robust aggregation rule \textsf{WFAgg}~\cite{cajaraville2025byzantine} to generate per-client alerts for malicious behavior. Each client weighs incoming neighbor model updates based on geometric and temporal statistics to determine trustworthiness, with a weight in the range of $[0, 1]$. The algorithm parameters are kept identical to the authors' selection, but the estimated malicious nodes are properly selected based on the specific topology. We validate the performance of the anomaly detection algorithm using the same policy to generate alert values, which will remain unchanged throughout the experiments for a fair comparison.

\textbf{Pre-training phase:} We enabled a pre-training phase of the FU-HST algorithm because it is an online streaming method, and we considered malicious behavior from the start of the DFL training. This pre-training phase is performed in a benign DFL setting with $15$ rounds, a $20$-node topology, and an $8$-regular communication graph, i.e., each node has exactly eight neighbors. Our goal is to enable the FU-HST algorithm to learn from outcomes of the aggregation rule \textsf{WFAgg} under a different topology (to avoid overfitting) and a stable topology (homogeneous number of neighbors) compared to those (random) considered during online training.

\textbf{FU-HST parameters:} Hyperparameter selection is applied to the underlying HST model parameters (number of trees $t$, tree depth $h$, anomaly threshold $\tau$, and window size $\psi$), using a simple grid search over the previous pre-training phase to identify a robust configuration. Concretely, we sweep $t\in [60,360]$ with step $60$, $h\in\{2,3,4,5,6\}$, $\tau\in[0.50,0.90]$ with step $0.05$, and $\psi\in[60,360]$ with step $60$, and select the setting that yields consistently high F1 across Noise and Sign-Flipping scenarios. The resulting configuration is $t=240$, $h=3$, $\tau=0.55$, and $\psi=120$, which we use in all experiments. The remaining FU-HST parameters are design-level controls rather than tuned hyperparameters and are kept fixed across all scenarios (to avoid confounding the evaluation). Specifically, the hysteresis lower threshold is set with $\gamma=0.5$ to allow for a wide range of suspicious activity and to avoid hasty decisions. Score-buffer EMA is set to $\alpha=0.5$ to react quickly to sustained deviations, while the feedback-buffer update uses $\beta=0.9$ for faster update when a client is flagged as malicious. Finally, stochastic update over non-flagged clients is set to $p_\text{u} = 0.05$ to regularize drift.

\textbf{Implementation details:} We perform validation tests using a Python simulator called \texttt{DecentralizedFedSim}~\cite{cajaraville2025byzantine}, which is enabled by the \texttt{PyTorch} library and models client behavior using threads. In addition to the baseline implementation, we modified the client workflow to send alerts after the aggregation phase and apply banning decisions. An object-based service models the round-loop process and applies the selected anomaly detection algorithm to the received client alerts.

\subsection{Comparison Baselines and Attack Models}
\label{subsec:compasion_baselines}

To evaluate and compare the performance of our proposal against other state-of-the-art algorithms, we have considered several anomaly detection algorithms:
\begin{itemize}
    \item \textbf{HST~\cite{tan2011fast}:} This is the standard HST algorithm that is considered in our FU-HST proposal. Despite maintaining all essential features as inputs, it does not implement the score stabilization method, dual-decision rule, or safe update. Hyperparameter selection is equal to those used for FU-HST algorithm.
    \item \textbf{SAD~\cite{hochenbaum2017automatic}:} ``Standard Absolute Deviation'' is a lightweight streaming detector that models normal behavior via location mean and a dispersion statistic. It assigns each new sample an anomaly score equal to its normalized absolute deviation from the current mean. In our setting, a node is classified as anomalous when its score exceeds a tuned threshold (set to $2.0$), consistent with commonly used SAD-style detectors.
    \item \textbf{iLOF~\cite{pokrajac2007incremental}:} ``Incremental Local Outlier Factor'' algorithm detects anomalies by comparing a point's local density to the densities of its $k$-nearest neighbors and updates neighborhood and density estimates via sliding-windows. Anomalies are declared by thresholding a score set to $1.13$ and neighborhood size to $75$, both held constant across all scenarios for fair comparison.
\end{itemize}

Conversely, several Byzantine attacks are considered to assess the performance and adaptability of the anomaly detection algorithms under different maliciousness scenarios. We assess the following state-of-the-art model poisoning attacks:
\begin{itemize}
    \item \textbf{Noise Attack:} Malicious model updates are modified by injecting Gaussian noise, $\mathcal{N}(\mu,\sigma^2)$, into each parameter of the local model. In our tests, both the mean and variance are set to $\mu=\sigma^2=0.1$.
    \item \textbf{Sign-Flipping:} Byzantine node reverses the direction of the model update without altering its norm.
    \item \textbf{Inner Product Manipulation:} IPM-$\varepsilon$ attack crafts a malicious update so that the aggregate update has a negative inner product with the true mean of the honest updates, i.e., the aggregation is pushed to point in the opposite direction of the intended descent direction. The parameter $\varepsilon$ governs the extent of malicious updates; for instance, we set $\varepsilon=100$ so the sign is reversed and the magnitude is increased.
\end{itemize}

\subsection{Validation Metrics}

Validation experiments are used to evaluate two distinct aspects: anomaly detection performance and the DFL results. For the former, given that it is a binary classification problem, we will use the F1 and accuracy metrics, which can be defined as
\begin{align}
    \mathsf{F1} &=  \dfrac{2 \times \mathsf{TP}}{2 \times \mathsf{TP} + \mathsf{FP} + \mathsf{FN}}, \\
    \mathsf{Accuracy} &= \dfrac{\mathsf{TP} + \mathsf{TN}}{\mathsf{TP} + \mathsf{TN} + \mathsf{FP} + \mathsf{FN}},
\end{align}
where $\mathsf{TP}, \mathsf{FP}, \mathsf{TN}, \mathsf{FN}$ are the standard confusion-matrix counts (true/false positives/negatives). Our primary focus will be on the F1 metric because we aim to evaluate both ``recall'' (identifying all malicious nodes), and ``precision'' (identifying a specific number of nodes correctly). Additionally, we utilize the false-ban rate, defined as the fraction of benign nodes incorrectly flagged as malicious:
\begin{equation}
    \mathsf{FBR} =  \dfrac{\mathsf{FP}}{\mathsf{FP} + \mathsf{TN}}
\end{equation}
For DFL model performance (considering the MNIST dataset), we use standard multi-class classification accuracy, i.e., the fraction of correct predictions over the test set, defined as
\begin{equation}
    \mathsf{Accuracy} = \dfrac{1}{|\mathcal{D}_\text{test}|} \sum_{i=1}^{|\mathcal{D}_\text{test}|} \mathbf{1}(\hat{y}_i = y_i),
\end{equation}
where $\mathcal{D}_\text{test}$ is the client-local testset, $(\mathbf{x}_i,y_i) \in \mathcal{D}_\text{test}$ denotes the $i$-th sample with feature vector $\mathbf{x}_i$ and label $y_i$, and $\hat{y}_i$ is the predicted label for $\mathbf{x}_i$.

\section{Results and Discussion}
\label{sec:results}

This section presents the validation results of the SDN-enabled anomaly detection mechanism under the experimental setup described in Section~\ref{sec:experimental_setup}, thereby assessing the selected DFL use case as a representative instantiation of the proposed architecture rather than as an exhaustive validation of all its architectural capabilities. We assess both (i) the anomaly classification performance and (ii) the resulting DFL learning performance, considering single-domain and multi-domain deployments to capture different topology conditions. Firstly, we evaluate FU-HST scalability in single-domain scenarios by varying the number of participants and the fraction of malicious nodes, and we compare it against the considered baselines (Subsection~\ref{subsec:anomaly_evaluation}). Next, we extend the evaluation to multi-domain scenarios to analyze robustness to inter-/intra-domain connectivity and to characterize performance under cross-domain neighbor exchanges (Subsection~\ref{subsec:multidomain_validation}). Finally, we analyze the computational complexity of FU-HST and quantify the computation and communication overhead introduced by SDN-enabled security enhancement relative to the baseline DFL execution (Subsection~\ref{subsec:overhead_analysis}).

\subsection{Anomaly Detection Evaluation}
\label{subsec:anomaly_evaluation}

\subsubsection{Scalability Analysis}

Firstly, we consider a set of single-domain DFL topologies by varying the number of nodes from $20$ to $100$. As stated in Section~\ref{sec:experimental_setup}, the communication graph is generated as an Erd\H{o}s--R\'enyi random graph $G(n,p_1)$. For each value of $N$, we select $p_1$ such that the expected node degree targets an average of $8$ neighbors per node, which is convenient to ensure stable behavior of the Byzantine-robust aggregation rule and, at the same time, enables a controlled scalability study as the topology becomes relatively sparser when $N$ increases. Moreover, the number of malicious nodes is fixed to $3$ in order to preserve a comparable malicious presence across all graph sizes while maintaining sufficient benign neighborhood for alert generation.

\begin{figure*}[tbp]
    \centering
    \includegraphics[width=1.0\linewidth]{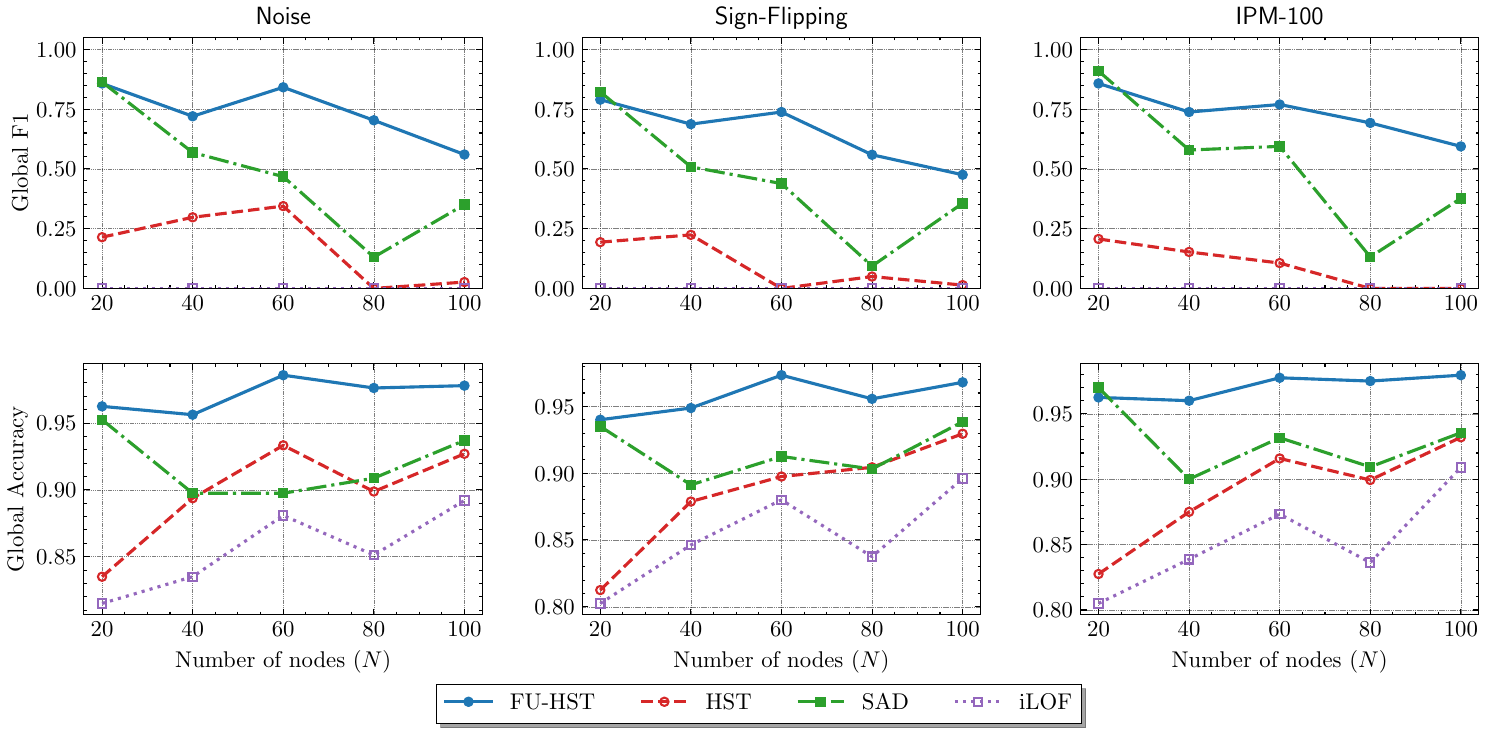}
    \caption{Comparison of the global F1 and Accuracy among all rounds in different DFL topologies with 3 malicious nodes and a range of 20 to 100 nodes.}
    \label{fig:f1_accuracy_comparison}
\end{figure*}

Figure~\ref{fig:f1_accuracy_comparison} shows the anomaly classification performance (F1 and accuracy) when sweeping the number of nodes. These metrics are computed by aggregating all banning decisions across all rounds (i.e., a single confusion matrix over the full execution), which jointly captures both the ability to identify malicious behavior and the number of rounds required for the detector to reach consistent decisions. Under this evaluation, F1 is expected to be lower than accuracy because accuracy is dominated by the majority benign class, whereas F1 directly penalizes missed detections and false alarms. Overall, considering the different Byzantine attacks, FU-HST achieves the best or near-best performance for most graph sizes compared to the considered baselines. In terms of F1, most methods degrade as $N$ increases (i.e., evidence per node becomes relatively more variable), whereas FU-HST exhibits a slower degradation trend, highlighting the robustness introduced by the score stabilization method. This effect is particularly visible when comparing FU-HST against the baseline HST for $N \ge 60$, where FU-HST maintains higher F1 as the topology becomes sparser. SAD performs competitively at $N=20$, but its performance does not generalize consistently as $N$ grows, suggesting sensitivity in alert statistics relative to the pre-training phase. Finally, accuracy remains high across most configurations because benign nodes constitute the majority of participants and are often correctly identified.

\begin{figure*}[tbp]
    \centering
    \includegraphics[width=1.0\linewidth]{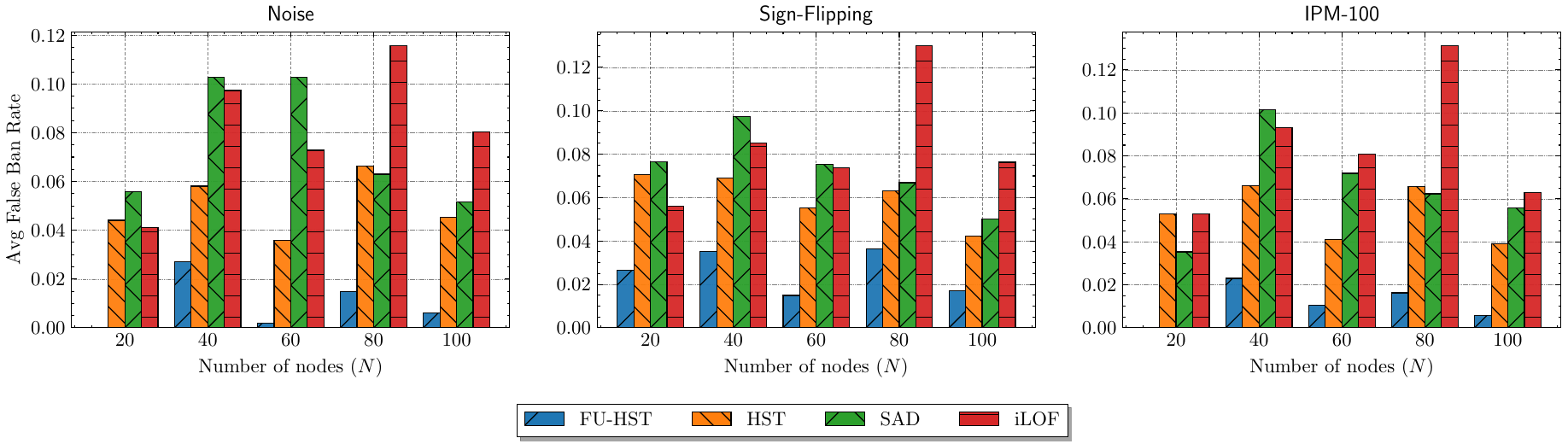}
    \caption{Comparison of the average false ban rate among all rounds in different DFL topologies with 3 malicious nodes and a range of 20 to 100 nodes.}
    \label{fig:false_ban_rate_comparison}
\end{figure*}

We also evaluate FBR as shown in Figure~\ref{fig:false_ban_rate_comparison}. Here, FBR is computed as the average across rounds of the fraction of benign nodes incorrectly flagged by the detector. FU-HST achieves the lowest FBR across all tested graph sizes and attacks, and notably does not exhibit an increasing false-ban trend as $N$ grows. This indicates that improved recall does not come at the expense of systematically over-banning benign participants.

\subsubsection{Threat Analysis}

\begin{figure*}[tbp]
    \centering
    \includegraphics[width=1.0\linewidth]{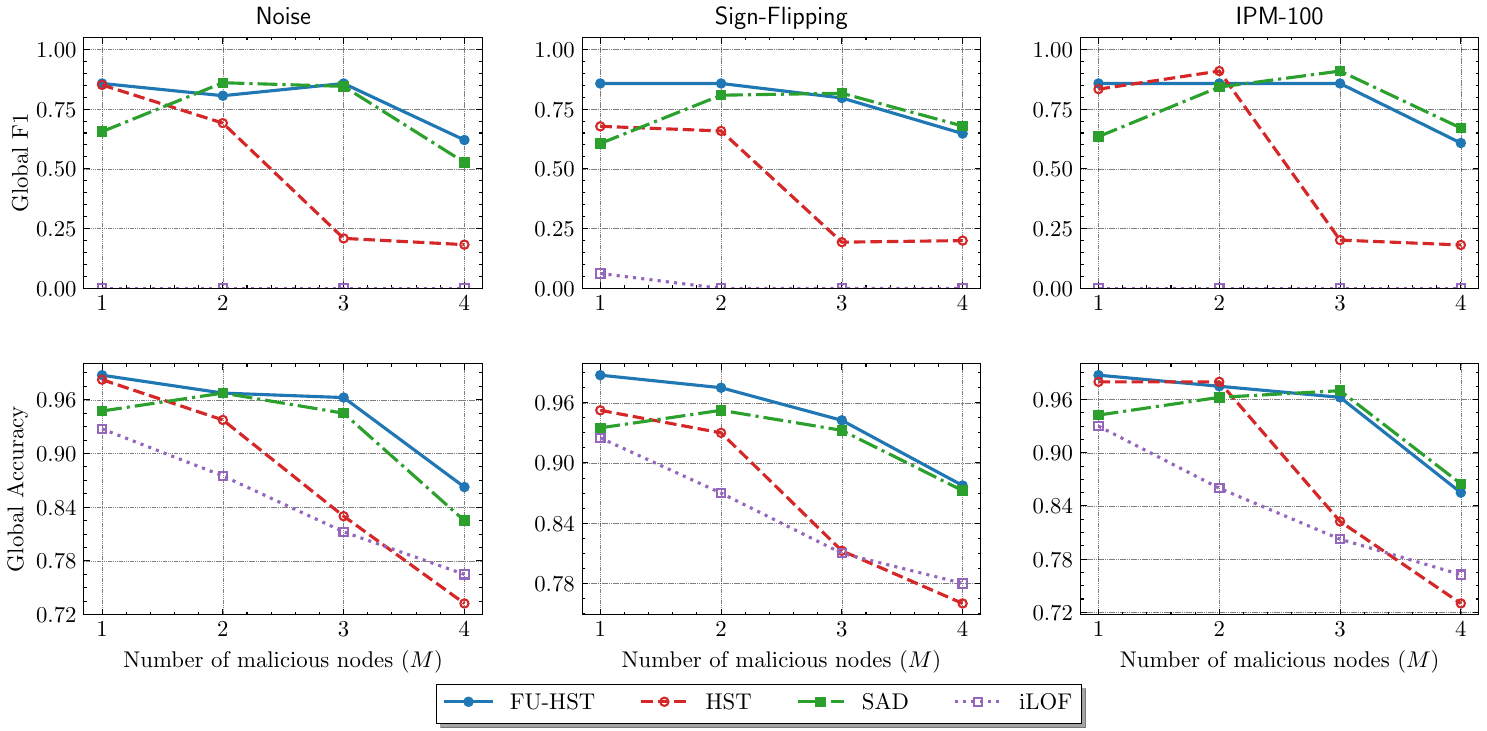}
    \caption{Comparison of the global F1 and Accuracy among all rounds in different DFL topologies with 20 nodes and a range of 1 to 4 malicious nodes.}
    \label{fig:f1_accuracy_malicious}
\end{figure*}

Conversely, we perform a similar validation test to assess the robustness of anomaly detection in terms of the number of malicious nodes. In this case, we consider a random topology with $20$ nodes and the same criterion for the probability $p_1$ (each node has $8$ neighbors, on average). During the experiments, the number of malicious nodes is swept from $1$ to $4$ in order to stay within the convergence regime of the aggregation rule, i.e., ensuring that benign updates remain the majority of the received ones during aggregation phase. As shown in Figure~\ref{fig:f1_accuracy_malicious}, FU-HST obtains the most stable F1 results when increasing the number of malicious nodes, with SAD showing slightly worse performance under this specific configuration. This can be justified since, as explained before, SAD behaves better when online statistics are evaluated under alert distributions close to the pre-training phase (here, the number of nodes is equal, while only the malicious fraction and topology varies). Clearly, FU-HST outperforms the baseline HST in both F1 and accuracy, where HST degrades quickly as the malicious fraction increases.

\begin{figure*}[tbp]
    \centering
    \includegraphics[width=1.0\linewidth]{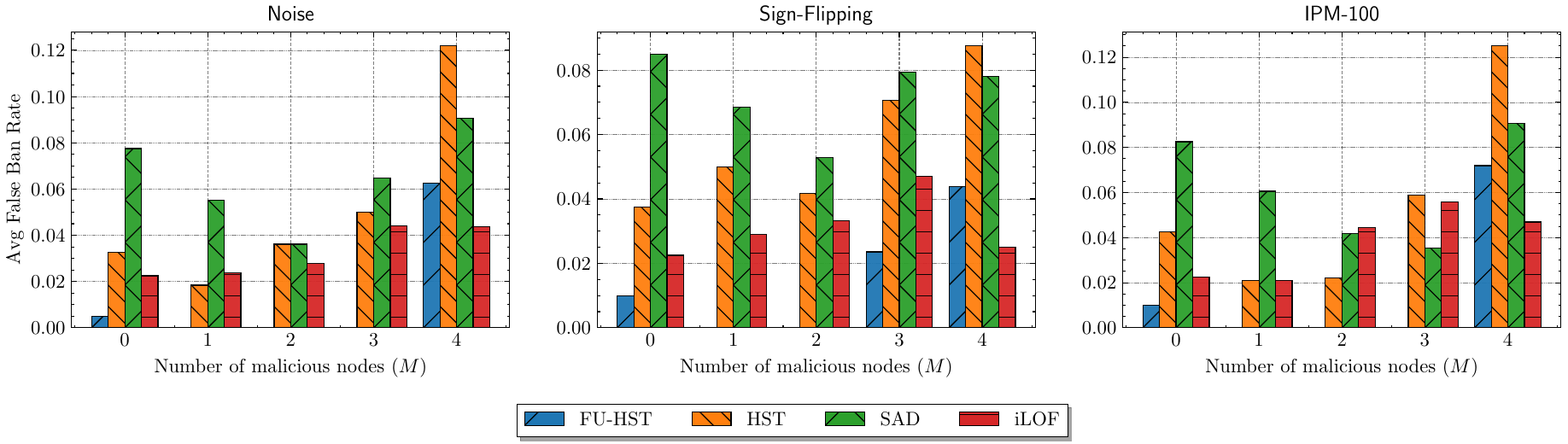}
    \caption{Comparison of the average false ban rate among all rounds in different DFL topologies with 20 nodes and a range of 0 to 4 malicious nodes.}
    \label{fig:false_ban_rate_malicious}
\end{figure*}

Additionally, similar to the scalability results, we evaluate the average FBR but when increasing the number of malicious nodes. As shown in Figure~\ref{fig:false_ban_rate_malicious}, FU-HST achieves the lowest FBR across all validation scenarios compared to the baselines. However, we observe that FBR increases for all algorithms when $M=4$, including our proposal. This behavior is expected because the local aggregation is close to its convergence limit (e.g., approximately $4$ malicious updates out of $8$ neighbors on average), which can bias the alert generation process and, consequently, lead to false positives during the anomaly detection.

\subsection{Multi-Domain Validation Scenarios}
\label{subsec:multidomain_validation}

Table~\ref{tab:multidomain_validation} evaluates the proposed FU-HST mitigation loop under a set of representative validation scenarios spanning both single-domain and multi-domain deployments. The evaluated scenarios, S1--S8, differ in the number of domains $D$, federation size $N$, malicious fraction $M$, inter-domain connectivity (captured by $p_2$ as defined in Section~\ref{subsec:validation-system}), malicious placement pattern, and the considered Byzantine attacks. For each scenario, we report DFL learning performance through test accuracy under three schemes: No Action (NA), where no mitigation is applied; Mitigation at rounds 10 and 20 (MIT@R10, MIT@R20), where FU-HST is enforced during training; and an Oracle baseline (ORA), which represents ideal filtering with perfect knowledge of malicious nodes. In addition, we report FU-HST detection performance through global F1-score and FBR. The goal of these validations is to verify that the mitigation mechanism either improves learning under Byzantine threats or does not degrade convergence while operating with bounded false banning (due to noisy alerts from nodes).

Scenarios S1--S3 consider a single domain with $N = 20$ and $M = 3$ malicious clients under the three considered attacks. In these settings, NA, MIT and ORA accuracy values are close, indicating that the underlying Byzantine-robust aggregation already provides a relatively stable baseline against these attacks. On the other hand, the mitigation mechanism remains non-disruptive, suggesting that FU-HST does not destabilize learning performance even when its benefit is limited by an already robust baseline. For instance, S1 shows a small gain at convergence comparing MIT@R20 and NA results ($0.859$ vs. $0.853$); S2 also improves slightly ($0.843$ vs $0.831$); and S3 mitigation is marginally below NA ($0.973$ vs. $0.977$) while remaining close to ORA ($0.976$). Finally, detection metrics remain moderate but consistent (global F1 around $0.30$) with bounded FBR (around $0.11$), indicating stable behavior in the single-domain regime. Although these values are worse than those observed in the previous detection-focused experiments, this is expected in this setting because FU-HST outputs are enforced as bans that modify the interaction graph. In particular, global F1 is computed over all rounds under the same malicious set; once a malicious node is banned, benign nodes stop generating alerts about banned nodes in subsequent rounds, while still counting the node as malicious in every round. At the same time, once a benign neighbor (i.e., a false ban) is excluded, it will remain filtered in subsequent aggregations, which increases its overall impact on the FBR. Therefore, detection metrics in this subsection reflect the performance of an online mitigation mechanism, rather than an offline classifier evaluated on a fixed stream.

\begin{table*}[tbp]
\centering
\caption{Single- and multi-domain performance validation under Byzantine attacks. DFL test accuracy is reported under No Action (NA), Mitigation at rounds 10 and 20 (MIT@R10 and MIT@R20), and Oracle (ORA) schemes. Detection performance is reported as global F1-score and false-ban rate (FBR).}
\label{tab:multidomain_validation}
\renewcommand{\arraystretch}{1.12}
\resizebox{\textwidth}{!}{%
\begin{tabular}{c c c c l l | c c c c | c c}
\toprule
\multicolumn{6}{c|}{\textbf{Scenario}} &
\multicolumn{4}{c|}{\textbf{DFL Accuracy}} &
\multicolumn{2}{c}{\textbf{Detection}} \\
\cmidrule(r){1-6}\cmidrule(lr){7-10}\cmidrule(lr){11-12}
\textbf{ID} & $D$ & $(N,M)$ & $p_2$ & \textbf{Malicious placement} & \textbf{Attack} &
\textbf{NA} & \textbf{MIT@R10} & \textbf{MIT@R20} & \textbf{ORA} &
\textbf{F1} & \textbf{FBR} \\
\midrule \midrule
S1  & 1 & (20,3) & -- & Random & Noise 
& 0.853 & 0.830$\pm$0.296 & 0.859$\pm$0.283 & 0.852 & 0.316 & 0.109 \\
S2  & 1 & (20,3) & -- & Random & IPM-100 
& 0.831 & 0.827$\pm$0.305 & 0.843$\pm$0.315 & 0.829 & 0.316 & 0.106 \\
S3  & 1 & (20,3) & -- & Random & Sign-Flipping 
& 0.977 & 0.954$\pm$0.015 & 0.973$\pm$0.013 & 0.976 & 0.293 & 0.118 \\
\midrule
S4  & 2 & (40,4) & 0.02 & Distributed & Sign-Flipping 
& 0.897 & 0.130$\pm$0.039 & \textbf{0.883$\pm$0.057} & 0.901 & 0.225 & 0.112 \\
S5  & 2 & (40,4) & 0.02 & Distributed & IPM-100 
& 0.101 & 0.126$\pm$0.031 & \textbf{0.677$\pm$0.249} & 0.721 & 0.261 & 0.140 \\
S6 & 3 & (40,3) & 0.03 & Inter-domain attacks & Sign-Flipping 
& 0.664 & 0.226$\pm$0.182 & 0.658$\pm$0.345 & 0.682 & 0.432 & 0.047 \\
S7 & 3 & (40,3) & 0.03 & Inter-domain attacks & IPM-100 
& 0.824 & 0.289$\pm$0.275 & 0.822$\pm$0.226 & 0.842 & \textbf{0.581} & \textbf{0.029} \\
S8 & 3 & (40,3) & 0.03 & No inter-domain attacks & IPM-100 
& 0.801 & 0.237$\pm$0.184 & 0.799$\pm$0.244 & 0.796 & 0.353 & 0.053 \\
\bottomrule
\end{tabular}%
}
\end{table*}

Scenarios S4--S8 extend the evaluation to multi-domain deployments with inter-domain links (parameterized by $p_2$) and different adversarial placement patterns. Here, the learning dynamics are more sensitive and converge less stably to topology changes because training depends on cross-domain neighbor exchanges, which are an inherent consequence of clustered multi-domain DFL graphs. In this regime, mitigation is expected to be most beneficial when poisoning is strong enough to disrupt learning beyond what Byzantine-robust aggregation can handle. Accordingly, a prominent improvement is observed in S5 (two domains, spread placement) under IPM-100 attacks where NA collapses in both domains ($0.101$), while mitigation achieves a substantial recovery ($0.677$), approaching the oracle baseline ($0.721$) with subsequent rounds. These results support one of the roles of FU-HST as a complementary mechanism to robust aggregation: when poisoning is strong enough to break the learning process, independent domain-side mitigation can restore a fraction of the attainable performance.

In contrast, in other multi-domain scenarios the final DFL accuracies are close (e.g., in S7 shows NA $0.824$ vs. MIT $0.822$; or S8 NA $0.801$ vs. MIT $0.799$), suggesting that the underlying aggregation rule already withstands the attack. In these cases, FU-HST remains non-disruptive while exhibiting good detection results under different placements of malicious nodes. For example, S7 (with inter-domain attacks) attains the best F1 detection quality ($0.581$) with the lowest false banning ($0.029$), which confirms the consistency of effective identification of Byzantine attacks that actively traverse domain boundaries; also S8 (with only intra-domain attacks, although their effects may still propagate across domain boundaries), which maintains low FBR ($0.053$) while achieving $\text{F1} = 0.353$, i.e., detectors still operate when malicious attacks are initially confined within a domain.

Overall, across several domain scenarios and adversarial placements, FU-HST mitigation does not prevent convergence and preserves low FBRs, while providing significant gains in scenarios where baseline learning degrades severely. Moreover, these results illustrate that validating security enhancement in multi-domain distributed learning is intrinsically challenging since performance depends jointly on attack strength, the placement of adversaries, and the topology-induced mixing dynamics of decentralized learning. In our analysis, the proposed evaluation aims to demonstrate that the SDN-enabled mechanisms can be deployed as a domain-side runtime application enhancement that remains stable across diverse multi-domain conditions, and that it provides clear benefit in the deployments where Byzantine threats exceed the robustness of aggregation rules.

\subsection{Complexity and Overhead Analysis}
\label{subsec:overhead_analysis}

We assess the practical cost of the proposed FU-HST security mechanism by combining both its theoretical per-round time and space complexity, and empirical computation and communication overhead measured in multi-domain DFL simulations. Following the analysis of work~\cite{tan2011fast}, processing one streaming instance through an ensemble of $t$ trees has average computational complexity $\mathcal{O} \left( t(h+1) \right)$, while the worst-case becomes $\mathcal{O} \left( t(h+\psi) \right)$ when window updates and resets occur between streaming instances, where $h$ is the maximum depth of trees and $\psi$ the considered window size. In our setting, each DFL round corresponds to scoring (and conditionally updating) one feature instance per participant in the domain. The additional data treatment introduced by our method (i.e., computing the feature vector) costs $\mathcal{O}(|\mathcal{R}_j^{(t)}|)$ per node. Hence, the per-round (average) computational complexity in domain $d$ is
\begin{equation*}
    \mathcal{O} \left( \sum_{j\in\mathcal{V}_d} |\mathcal{R}_j^{(t)}| + |\mathcal{V}_d|t(h+1) \right)
\end{equation*}
On the other hand, HST space complexity is $\mathcal{O}\left( t 2^h \right)$, which is constant for fixed $t$ and $h$ values. Additional per-node buffers $(\mathbf{s},\mathbf{f},\mathbf{c})$ contribute $\mathcal{O}(|\mathcal{V}_d|)$ space; therefore, the overall space complexity in domain $d$ is
\begin{equation*}
    \mathcal{O} \left( t 2^h + |\mathcal{V}_d| \right).
\end{equation*}
Since this space analysis in memory terms is determined by fixed HST parameters and buffers, we focus our experimental evaluation on time and bandwidth overhead.

\begin{figure}[tbp]
    \centering
    \includegraphics[width=0.52\linewidth]{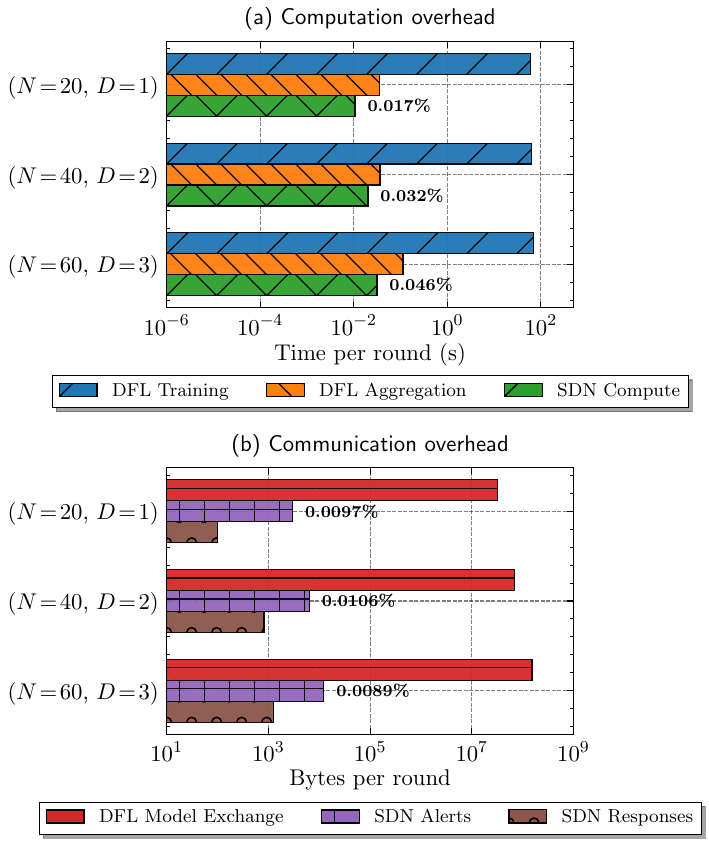}
    \caption{Computation and communication overhead of the SDN-enabled anomaly detection loop in multi-domain DFL process.}
    \label{fig:overhead_analysis}
\end{figure}

Figure~\ref{fig:overhead_analysis} analyzes the additional overhead, both computation and communication, introduced by the SDN-enabled anomaly detection mechanism relative to the underlying DFL process. To assess this validation, we consider three topology scenario of increasing multi-domain complexity defined by the following pairs: $(N=20,D=1), (N=40,D=2), (N=60,D=3)$. Intra-domain connectivity is configured to maintain an average degree of $8$ (as in previous testing) and the inter-domain connectivity is adjusted to increase proportionally the cross-domain communication with $p_2$ taking values $0.0$, $0.08$ and $0.15$ in each scenario, respectively. In all scenarios it is kept a fraction of malicious nodes of $10\%$, all of them applying Byzantine attack Sign-Flipping (just a use case since it does not affect the overhead analysis). Reported values correspond to mean computation time per round (computation subplot) and mean transmitted bytes per round (communication subplot).

The FU-HST computation overhead is consistently low across all three scenarios. In first scenario, the anomaly detection requires, on average, $0.0106$ seconds per round, which corresponds to $0.0175 \%$ of the DFL process time (both training and aggregation phases). In the second and third scenarios, SDN computation increases to $0.0201$ and $0.0313$ seconds per round, respectively, while remaining below $0.05 \%$ of DFL process time. These results are expected as the FU-HST processes a larger number of alerts when $N$ and $D$ increase, specially with the inter-domain probability $p_2$. However, the dominant cost remains local model training, and the FU-HST computation overhead stays negligible even under higher inter-domain connectivity. Notably, FU-HST computation remains also smaller than DFL aggregation time in all cases, confirming that the proposed security enhancement mechanism does not alter the computational bottleneck of the DFL system.

The communication overhead is even lower between DFL process and the SDN-enabled security enhancement mechanism. Average per-round model exchange ranges from $31.7$\,MB (in the first single-domain scenario) to $151.6$\,MB (in the third scenario), whereas SDN alerts and responses remain in the order of kilobytes. For example, $2,981 + 89$\,bytes in the single-domain scenario and $12,274 + 1,238$\,bytes in third one. Consequently, the relative communication overhead stays around $0.01\%$ across all scenarios, despite the growth in network size and inter-domain links. These results indicate that the FU-HST algorithm is suitable for multi-domain DFL deployments with minimal additional compute and bandwidth requirements compared to the underlying DFL deployment.

\section{Conclusions and Future Work}
\label{sec:conclusions}

This paper introduced a multi-domain architecture for orchestrating fluid computing environments where distributed applications span heterogeneous resources owned by independent domains. The proposed framework is primarily based on a decentralized per-domain orchestrator as its primary capability that, in practice, generates an agnostic and unified fluid computing platform for deploying tenants' applications. In other words, tenants only express intent-based deployment requests through management endpoints, while each domain enforces local autonomy and enables decentralized coordination to ensure cross-domain consistency during runtime execution. This enables the framework to facilitate E2E continuity of applications while maintaining explicit and realistic administrative boundaries. We also instantiated the architecture with a representative distributed AI workload to validate that domain-side control services can be used for application-level enhancement. Specifically, we proposed an SDN-enabled security enhancement mechanism for multi-domain DFL deployments, based on per-domain anomaly detection techniques (that complement Byzantine-robust aggregation) and multi-domain coordination.

Evaluation tests validated the proposed use case across different federation sizes, malicious fractions, and domain configurations, including multi-domain deployments with diverse inter-domain connectivity and adversarial placements. The results show that the proposed SDN-enabled FU-HST mechanism can be deployed as a domain-side runtime enforcement loop for DFL, providing consistent mitigation behavior across the evaluated Byzantine threats. We additionally quantified computation and communication overhead, showing that the added SDN-enabled security loop incurs limited per-round cost relative to the underlying DFL training process.

However, further studies are needed to enhance fluid computing environments. On the orchestration plane, an important next step is to formalize the behavior of the Multi-domain Coordination Agent within the Domain Service Orchestrator by defining and evaluating decentralized coordination algorithms under realistic domain constraints (limited visibility, bounded disclosure, and asymmetric capabilities), including game-theoretic formulations for negotiation and incentive-compatible resource/relocation agreements. Additionally, future work will also develop and evaluate concrete algorithmic solutions for fluid runtime adaptation. Specifically, offloading and migration policies will be designed to operate in distributed execution settings, including self-organized D2D clusters in 6G-like environments, treating them as first-class execution domains under policy-constrained offloading decisions.

\section*{Acknowledgment}
This work was supported by the grant PID2023-148716OB-C31 funded by MCIU/AEI/10.13039/501100011033 (DISCOVERY project); and ``TRUFFLES: TRUsted Framework for Federated LEarning Systems'', within the strategic cybersecurity projects (INCIBE, Spain), funded by the Recovery, Transformation and Resilience Plan (European Union, Next Generation). Additionally, it has also been funded by the Galician Regional Government under project ED431B 2024/41 (GPC). Finally, this research project was made possible through the access granted by the Galician Supercomputing Center (CESGA) to its supercomputing infrastructure.


\end{document}